\documentclass[pra,letterpaper,onecolumn,showpacs,superscriptaddress,floatfix]{revtex4}
\usepackage{graphicx,psfrag,amsmath,amssymb,amsfonts,bbm,latexsym,color,dcolumn,bm}

\begin{document}
\title{Thermal and dissipative effects in Casimir physics}

\author{M. Brown-Hayes}
\affiliation{Department of Physics and Astronomy,Dartmouth College,6127 Wilder Laboratory,Hanover,NH 03755,USA}

\author{J.H. Brownell}
\affiliation{Department of Physics and Astronomy,Dartmouth College,6127 Wilder Laboratory,Hanover,NH 03755,USA}

\author{D.A.R. Dalvit} 
\affiliation{Theoretical Division, MS B213, Los Alamos National Laboratory, Los Alamos, NM 87545, USA}

\author{W.-J. Kim}
\affiliation{Department of Physics and Astronomy,Dartmouth College,6127 Wilder Laboratory,Hanover,NH 03755,USA}

\author{A. Lambrecht} 
\affiliation{Laboratoire Kastler-Brossel, Universit\'e Pierre et Marie Curie, Campus 
Jussieu, F-75252 Paris, France}

\author{F.C. Lombardo}
\affiliation{Departamento de Fisica J.J. Giambiagi, Facultad de Ciencias Exactas y Naturales, 
Ciudad Universitaria, Pabell\'on 1, Universidad de Buenos Aires, 1428 Buenos Aires, Argentina} 

\author{F.D. Mazzitelli}
\affiliation{Departamento de Fisica J.J. Giambiagi, Facultad de Ciencias Exactas y Naturales, 
Ciudad Universitaria, Pabell\'on 1, Universidad de Buenos Aires, 1428 Buenos Aires, Argentina} 

\author{S.M. Middleman} 
\affiliation{Department of Physics and Astronomy,Dartmouth College,6127 Wilder Laboratory,Hanover,NH 03755,USA}

\author{V.V. Nesvizhevsky}
\affiliation{Institute Laue-Langevin, 6 rue Jules Horowitz, F-38042 Grenoble, France}

\author{R. Onofrio}
\affiliation{Department of Physics and Astronomy,Dartmouth College,6127 Wilder Laboratory,Hanover,NH 03755,USA}
\affiliation{Dipartimento di Fisica ``G. Galilei'', Universit\`a di Padova, 
Via Marzolo 8, Padova 35131, Italy}

\author{S. Reynaud}
\affiliation{Laboratoire Kastler-Brossel, Universit\'e Pierre et Marie Curie, Campus 
Jussieu, F-75252 Paris, France}

\begin{abstract}
We report on current efforts to detect the thermal and dissipative contributions 
to the Casimir force. For the thermal component, two experiments are in progress  
at Dartmouth and at the Institute Laue Langevin in Grenoble. 
The first experiment will seek to detect the Casimir force at the 
largest explorable distance using a cylinder-plane geometry which offers various 
advantages with respect to both sphere-plane and parallel plane geometries.
In the second experiment, the Casimir force in the parallel plane 
configuration is measured with a dedicated torsional balance, up to 10 $\mu$m. 
Parallelism of large surfaces, critical for this configuration, 
is maintained through the use of inclinometer technology already implemented at Grenoble 
for the study of gravitationally bound states of ultracold neutrons. 
For the dissipative component of the Casimir force, we discuss detection 
techniques based upon the use of hyperfine spectroscopy of ultracold atoms and 
Rydberg atoms. 
Although quite challenging, this triad of experimental efforts, if 
successful, will give us a better knowledge of the interplay between 
quantum and thermal fluctuations of the electromagnetic field and of 
the nature of dissipation induced by motion of objects in quantum vacuum.  
\end{abstract}


\maketitle

\section{Introduction}

The study of quantum vacuum \cite{Milonni} is of great interest in physics due to 
recent astrophysical observations supporting an accelerating universe \cite{Riess,Garnavich,Perlmutter}. 
In various models, the acceleration is attributable to a cosmological term 
triggered by quantum vacuum effects \cite{Peebles}. 
In addition, the development of models unifying gravity and the other interactions 
has led to predictions of the existence of new Yukawian forces in the micrometer 
range with coupling of the same order of magnitude as gravity \cite{Fischbach}.
These motivations are encouraging many theoretical and experimental studies into the 
physics of quantum vacuum and its interplay with cosmology and 
elementary particle physics \cite{Zeldovich,Weinberg,Carroll,Sahni}. 
The Casimir force \cite{Casimir} is an important accessible window in this context 
(see \cite{Plunienrev,Elizalde,Mostepanenkobook,Bordagbook,Bordagreview,Reynaudreview,Miltonbook,Miltonreview,Lamoreauxreview} 
for monographs and reviews) and, particularly in the last decade, has been 
studied with an increasing level of accuracy in various geometries, from 
parallel plates \cite{Sparnaay,Bressi} to sphere-plane 
\cite{vanblokland,Lamoreaux,Mohideen,Harris,Chan1,Chan2,Decca1,Iannuzzi,Decca2,Decca3}. 
Detailed knowledge of the Casimir force is important to master all the corrections in order to constrain, 
from the residuals of the actual experiments, the presence of Yukawian forces in 
the micrometer range. 
In particular, the interplay of Casimir forces with thermal photons due to 
a blackbody background and in presence of realistic 
cavities is not fully understood, and it seems likely that only experiments 
will be able to discern among the models proposed so far. 
The contribution of thermal photons grows with the size of the cavity in 
which the Casimir pressure is exerted, thus one needs to measure 
the Casimir force at large distances. Given the reduced sensitivity of 
atomic force microscope experiments at large distances and the weak 
signal expected with a sphere-plane geometry, the most promising 
configurations to look for Casimir forces at large distances are 
the cylinder-plane and the parallel plane geometries.

Thermal effects are not solely a result of finite temperature environment, 
since the motion of objects in quantum vacuum in itself also gives rise 
to heating. In fact, it has been predicted that non uniformly accelerated objects 
should dissipate energy in the form of photons, a phenomenon known as the dynamical 
Casimir effect \cite{Moore,Law,Schwinger,Cole,Lambrecht,Plunien,Dodonov,Crocce}. 
When specialized in a cavity configuration, this effect is also equivalent to parametric production 
of photons through quantum vacuum. The predicted Casimir photon production is quite 
low for realistic configurations, necessitating low-noise, high-sensitivity detection 
techniques such as hyperfine or Rydberg atomic spectroscopy.
In this paper we discuss the current status of three ongoing experimental 
projects aimed at measuring the thermal contribution to the Casimir force and 
observing the predicted vacuum photons emitted from an non-uniformly accelerated 
resonant cavity. In section 2 we discuss the predictions for the thermal contribution 
to the Casimir force expected for both the cylinder-plane and the parallel plane configurations. 
We then describe some experimental issues for the corresponding experimental efforts 
ongoing at Dartmouth and in Grenoble, focusing on potential hurdles for the projects. 
Finally, in section 3, we briefly describe a proposal to detect the dynamical Casimir 
photons based upon atomic spectroscopy and high frequency mechanical resonators. 

\section{Thermal contribution to the Casimir force}

The thermal contribution to the Casimir force has not yet been detected, 
in spite of its importance as a background to new forces in the 
1 $\div$ 10 $\mu$m range. This scenario is complicated by the 
presence of models which yield different results when finite 
temperature and finite conductivity are taken into 
account  
\cite{Lambrecht2,Bostrom,Lamoreauxcomm,Serneliusreply,Chen,Geyer1,Geyer2,Esquivel,Genet1,Genet2,Svetovoy,Torgenson,Esquivel1,Bordag,Hoye,Brevik}. 
The basic theoretical formalism for the calculation of Casimir forces  
between real metals at finite temperature is given by the Lifshitz theory \cite{Lifshitz1,Lifshitz2}. 
This theory provides an expression for the pressure $P_{pp}$ (or, equivalently, the 
force $F_{pp} = S \;  P_{pp}$) between two parallel plates of surface area $S$, 
separated by a gap $d$,

\begin{equation}
F_{pp}(d)= \frac{S}{\pi \beta d^3} \sum_{m=0}^{\infty \;  
'} \int_{m \gamma(d)}^{\infty} dy \; y^2
\; \left[  \frac{r_{\rm TM}^{-2} e^{-2y}}{1-r_{\rm TM}^{-2} e^{-2y}}  +
\frac{r_{\rm TE}^{-2} e^{-2y}}{1-r_{\rm TE}^{-2} e^{-2y}} \right] .
\label{forcepp}
\end{equation}
Here  $\beta=1/k_{\rm B} T$ is the inverse temperature, 
$\gamma(d)=2 \pi d / \beta \hbar c$, and the prime on the summation 
sign indicates that the $m=0$ term is counted with half weight. 
The reflection coefficients $r_{\rm TE}$ and $r_{\rm TM}$ for the 
two independent polarizations TE and TM are computed at imaginary 
frequencies $\omega_m=i \xi_m$, where $\xi_m=2 \pi m / \beta \hbar$ 
are the Matsubara frequencies.

The exact expressions for the reflectivity coefficients, which   
encompass the optical response of the electrons in the metallic  
surfaces, are not known. Within the Lifshitz formalism, they are 
expressed in terms of the dielectric permittivity  $\epsilon(\omega)$
of the metals

\begin{eqnarray}
r_{\rm TM}^{-2} = \left[ \frac{\epsilon(i \xi_m) p_m + s_m}{\epsilon 
(i \xi_m) p_m - s_m} \right]^2  &;&
r_{\rm TE}^{-2} =\left[ \frac{s_m + p_m}{s_m - p_m} \right]^2 ,
\end{eqnarray}
where $p_m=y/ m \gamma$ and $s_m = \sqrt{\epsilon(i  \xi_m) - 1 + p_m^2}$. 
The dielectric permittivity along the imaginary frequency axis can be computed 
using tabulated optical data for different metals. For the range of temperatures 
to be probed in the experiments (a few degrees around $T=300$ K), permittivity 
data corresponding to Matsubara frequencies $\zeta_m=2 \pi m / \beta \hbar$ with 
$m \ge 1$  can be extracted from the optical data. However, to determine the $m=0$ contribution, 
it is necessary to extrapolate the data to zero frequency. This extrapolation has been done 
by several groups using different approaches, leading to contradicting predictions for the 
magnitude of the force.

\begin{figure}
\setlength{\unitlength}{1cm}
\begin{center}
\includegraphics*[width=7.5cm,angle=0]{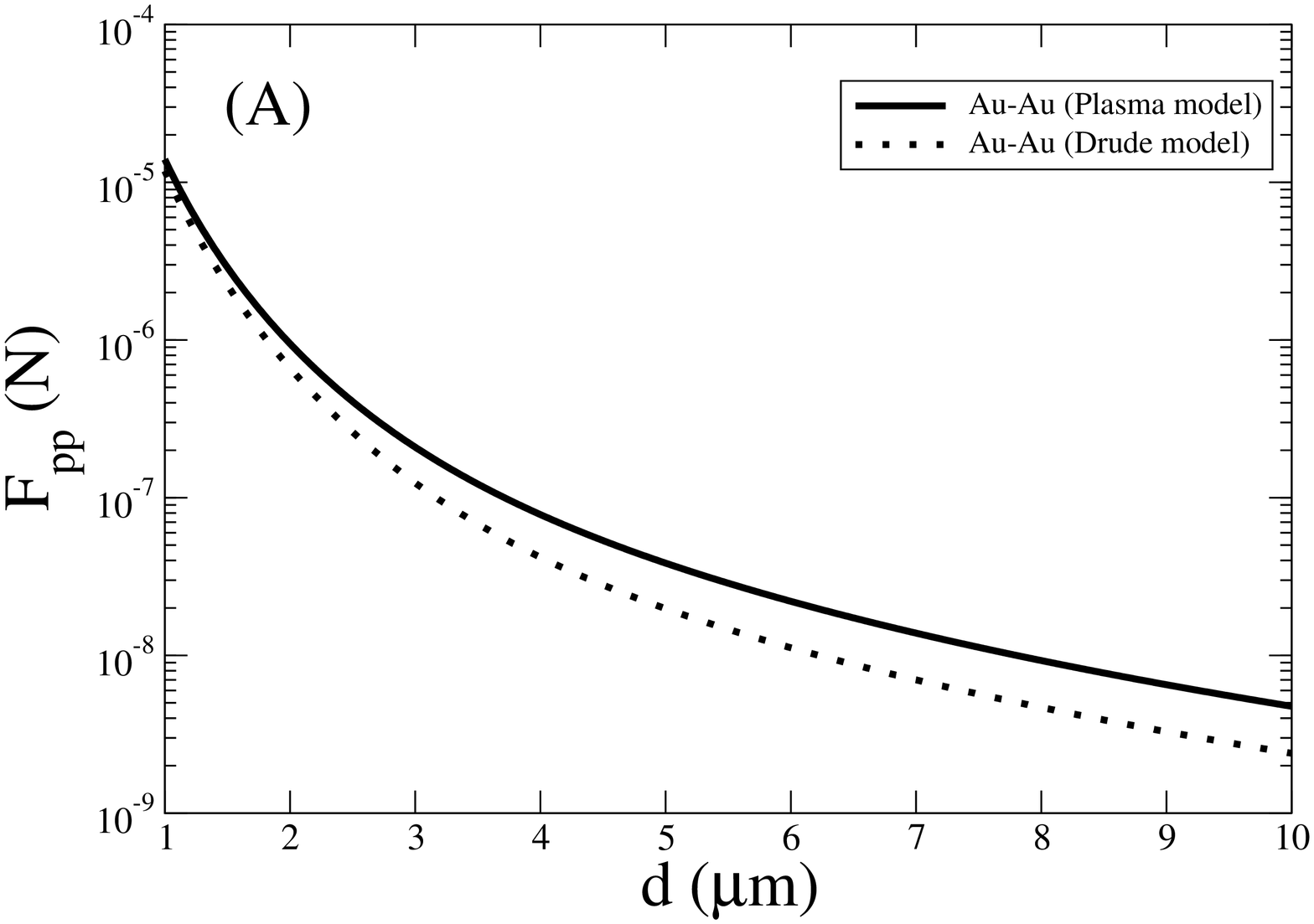}
\includegraphics*[width=7.5cm,angle=0]{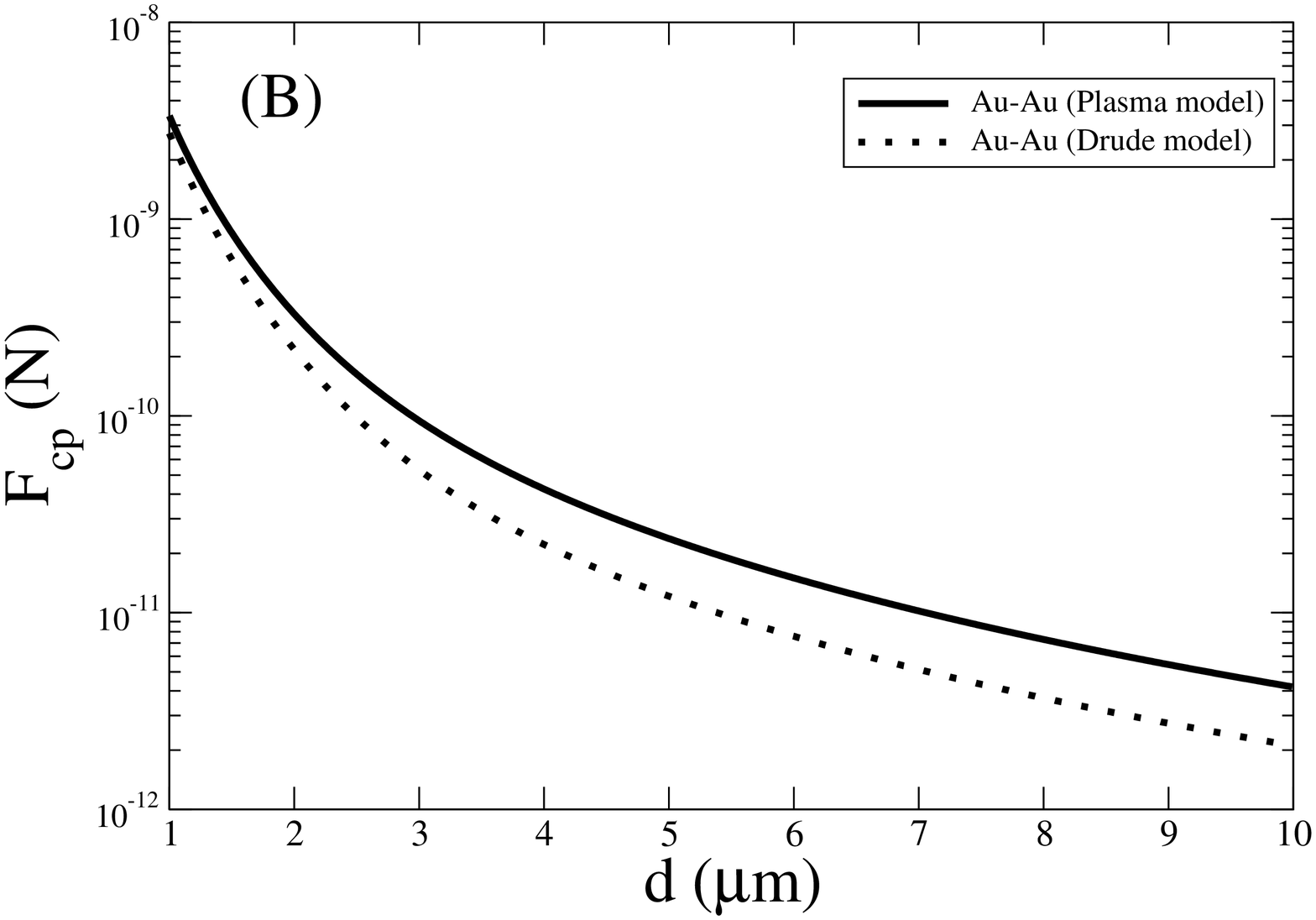}
\end{center}
\caption{Absolute Casimir force as a function of distance at $T=300$ K 
for Au-Au metallic surfaces in (A) plane-plane configuration  and (B) cylinder-plane
configuration. The force is computed using two distinct theoretical  approaches:
the plasma model (for which the TE $m=0$ mode contributes to the  
force), and the Drude model (for which the
TE $m=0$ mode does not contribute). The plasma frequency of Au is $ 
\omega_p=9.0 \; {\rm eV}$, and its relaxation frequency is
$\nu = 35 \; {\rm meV}$. For (A) we use $10 \; {\rm cm} \times 12 \;  
{\rm cm}$ plates. For (B) we use $L=2 \ {\rm cm}$ for the length of the  
cylinder, and $a=1\ {\rm cm}$ for its radius. In the range of distances  
plotted in these figures, the Casimir force between Cu-Cu and Al-Al  
surfaces is the same as for Au-Au surfaces within $0.1 \%$.
}
\label{fig1}
\end{figure}
In Fig. 1 (left) we show the absolute value of the force between two  
parallel gold metallic plates  at $T=300$ K
as a function of the gap using two different theoretical approaches.
The first approach is based on the plasma model 
$\epsilon(i \xi) = 1 + \omega_p^2/ \xi^2$, where $\omega_p$ is 
the plasma frequency. In this case,  the TE $m=0$ term contributes 
to the Casimir force. The second approach is based on the Drude model 
$\epsilon(i \xi)= 1 + \omega_p^2/ (\xi (\xi + \nu))$, where $\nu$ is the relaxation
frequency. In this case the TE $m=0$ term does not contribute to the Casimir force.
We have numerically computed the integral in Eq.(\ref{forcepp}) using quadratures 
with a cut-off $y_{\rm max}=50 + m \gamma$, and we have used
a cut-off $m_{\rm max}$ for the $m$ summation, corresponding to a  
maximum Matsubara frequency $\zeta_{\rm max} = 10^{17} {\rm rad}/{\rm  
sec}$.
We have checked that the results are robust against variations of these  
cut-offs.
We also show in Fig. 1 (right) the force between a cylinder (of radius $a$ and length $L$) 
and a plane, assumed to be  parallel, separated by a gap $d$. 
This force has been evaluated using the  results for the  pressure in the parallel plane configuration, and  
the proximity force approximation (valid in the limit $d \ll a$) 
\cite{Derjaguin1,Derjaguin2,Blocki}.

\begin{equation}
F_{cp} =  2\int_0^{\pi/2}\,\, P_{pp}(d+a(1-\cos\varphi))L a
d\varphi\,\, . \label{fcpnew}
\end{equation}
The angle $\varphi$ parameterizes the location of the
infinitesimal surfaces on the cylinder.
The leading contribution to the integral comes from angles close
to $ \varphi =0$. In order to speed up the numerical computation
of the integral, we have used as an upper limit $\varphi=\pi/20$,
and we have checked that the results, in the limit $d \ll a$, are
robust when we varied this upper limit.

In the proximity force approximation the surfaces of the plane and
the cylinder are divided into infinitesimal parts to integrate the
parallel plates result. There is an ambiguity in the choice of the
areas of these infinitesimal parts. Different choices give
distinct approximations for the force. In Eq. \ref{fcpnew} above,
we have chosen the area of a small portion of the cylinder
$dA_c=Lad\varphi$. One could also use the area of a small portion
of plane $dA_p=La\cos\varphi d\varphi$, or a combination of the
two, like the geometric mean $dA_{gm}=(dA_pdA_c)^{1/2}$. The
accuracy of the proximity force approximation has been estimated
by computing the force $F_{cp}$ using all the areas mentioned
above. The results differ by less than $1.2\%$ for both plasma and
Drude models at $d<10\mu$m.

Analytical expressions for the force, both for parallel plates and  
cylinder-plate geometries,  can be obtained for some limits. We  
consider here the case of surfaces without roughness (corrections due  
to roughness are important for small distances $d < 1 \mu{\rm m}$ at  
$T=300$ K).
For perfect conductors at zero temperature one recovers the original  
Casimir formula for the parallel plates case, and the cylinder-plane
force can be obtained in the proximity force approximation ($d \ll a$) \cite{Dalvitepl}

\begin{equation}
F_{pp}^{(T=0)}= \frac{\pi^2 \hbar c }{ 240} \; \frac{S}{d^4} \; \; ;  
\; \;  F_{cp}^{(T=0)} = \frac{\pi^3 \hbar c L a^{1/2}}
{384 \sqrt{2} d^{7/2}} .
\end{equation}
The scaling of the force with distance for the cylinder-plane case ($ 
\simeq d^{-7/2}$) is intermediate between the
parallel plane ($\simeq d^{-4}$) and the sphere-plane ($\simeq d^{-3} 
$) cases. This is an advantage with respect to the parallel plane case 
because the latter is quickly going to zero with the distance.
For reasonable values of the relevant parameters, the magnitude 
of the Casimir force in the cylinder-plane case is also intermediate
between the cases of the sphere-plane, as investigated with atomic force
microscopy, and the case of parallel planes \cite{note}.
For finite temperature and real conductors, one can obtain closed  
expressions for the force using, e.g., the plasma model.
At temperatures $T \ll T_{\rm eff} = \hbar c / 2 k_{\rm B} d$,  
and to first order in $\delta/d$ (where $\delta = \lambda_p / 2 \pi$,  
with $\lambda_p$ the plasma wavelength) one obtains

\begin{eqnarray}
\frac{F_{pp}}{F_{pp}^{(T=0)}}  &\approx& 1 +  \frac{1}{3} \;   
\left(\frac{T}{T_{\rm eff}}\right)^{4} -
\frac{16 \delta}{3 d} \; \left[  1 - \frac{45 \xi(3)}{8 \pi^3} \;     
\left(\frac{T}{T_{\rm eff}}\right)^{3} \right]  \nonumber \\
\frac{F_{cp}}{F_{cp}^{(T=0)}}   &\approx& 1 + \frac{132.096  
\zeta(7/2)}{\pi^{9/2}} \; \left(\frac{T}{T_{\rm eff}}\right)^{7/2} -
\frac{\delta}{d} \; \left[  \frac{14}{3} - \frac{48 \xi(3)}{\pi^3}  
\;    \left(\frac{T}{T_{\rm eff}}\right)^{3} \right] , \nonumber
\end{eqnarray}
where $\xi(z)$ is the Riemann zeta function. Finally, in the high  
temperature limit $T \gg T_{\rm eff}$ the forces
are dominated by the thermal contribution

\begin{equation}
F_{pp}^{\rm thermal} \approx \frac{\xi(3) k_{\rm B} T }{4 \pi} \;  
\frac{S}{d^3} \;\; ; \;\;
F_{cp}^{\rm thermal} \approx \frac{3 \xi(3) k_{\rm B} T}{16 \sqrt{2}}  
\;  \frac{L a^{1/2}}{d^{5/2}} ,
\end{equation}

\noindent
which represent the radiation pressure due to the thermal blackbody 
photons at finite temperature. This force could be used, by intentionally 
increasing the temperature by known amounts, to obtain physical calibrations 
and estimates of the sensitivity for any apparatus aimed at high-precision 
measurements of the Casimir force.

\begin{figure}[b]
\setlength{\unitlength}{1cm}
\begin{center}
\includegraphics*[width=8.0cm,angle=0]{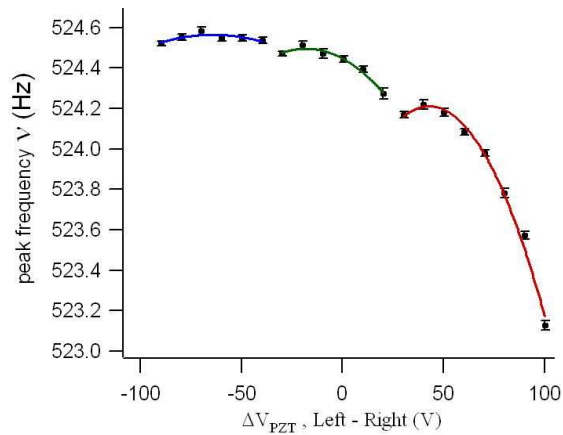}
\end{center}
\caption{Long-term behaviour of the resonator frequency.
Depicted is the frequency of the resonator with an applied 
bias voltage of $V=100$ V versus the difference between 
the voltages applied to the piezoelectric actuators, starting 
from the highest voltage $\mathrm{V_L-V_R}$=+100 V and scaling 
down in time (from right to left) to minimize hysteresis effects 
in the piezoelectric actuators. 
The free resonator frequency (without bias voltage) is 
$\nu_0=(526.40\pm 0.02)$ Hz.
Each measurement carries 100 averages (each average lasting 32 s), and 
after the first 7 measurements we experienced an abrupt change in distance 
(manifested both by a swallower curvature and a higher frequency) as well 
as in the angle formed by the cylinder axis and the resonator surface 
(manifested by a displacement in the voltage for which there is 
the minimum frequency shift).  
A second abrupt change has taken place after another 6 measurements, in 
the same direction for the drift of both the gap distance and angle. 
The continuous curves are the distinct parabolic fits to each set of data. 
The overall duration of the run was $\simeq$ 18 hours.}
\label{fig2}
\end{figure}

\subsection{Cylinder-plane configuration: experimental issues}

The apparatus we are developing to measure the Casimir force 
in a cylinder-plane configuration has been described in \cite{Brown} 
as are parallelization and calibration techniques. 
Here we focus on possible complications in measuring the Casimir force
due to the finite thermal stabilization of the apparatus. 
This is reflected first of all in the stability of the parallelism.
Long term thermal drifts on the apparatus affect the
parallelism and therefore limit the total measurement time.
As discussed in \cite{Brown}, the parallelism can be assessed by intentionally 
rotating the cylinder around its midpoint. The parallel configuration
is that which minimizes the frequency-shift due to an external 
spatially dependent force, such as the electrostatic force, and 
this is shown in Fig. 2 for a relatively long measurement time. 
The curves are fitted as pieces of parabolas but both the curvature 
and the center of the parabolas are different, with the shallower 
curve corresponding to a larger distance between the cylinder and the plane.
This can be interpreted as due to sudden drifts in the separation distance, and 
will be mitigated in future versions of the apparatus with improved 
thermal stabilization and the use of materials like Invar alloys 
with minimal thermal expansion coefficients.

\begin{figure}
\setlength{\unitlength}{1cm}
\begin{center}
\includegraphics*[width=7.0cm,angle=0]{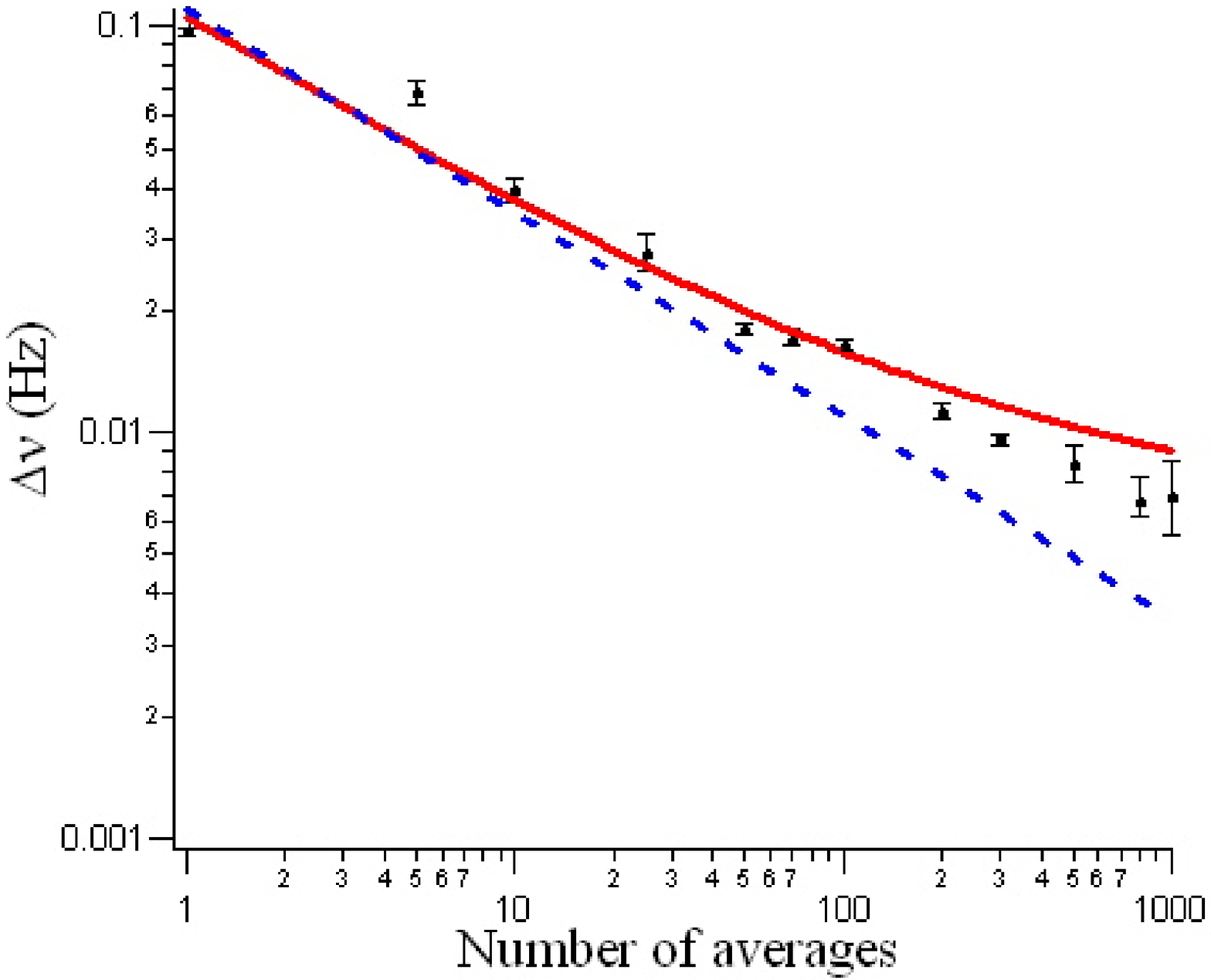}
\includegraphics*[width=7.0cm,angle=0]{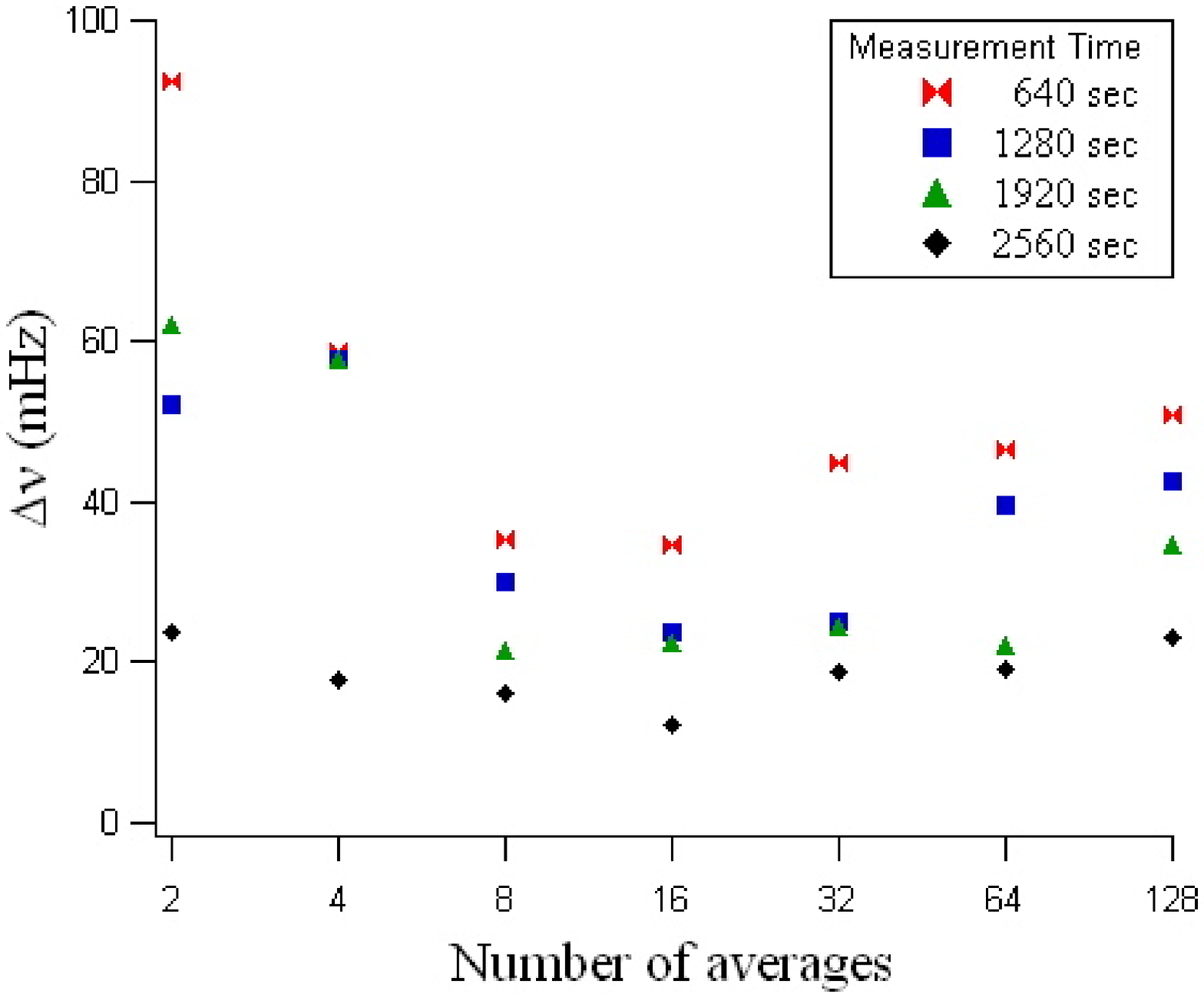}
\end{center}
\caption{Optimization of the accuracy for the peak frequency measurement.
On the left panel  we plot the peak frequency uncertainty  
evaluated from the fitting of the resonance curve of the mechanical 
oscillator versus the number of averages taken with the FFT 
spectrum analyzer at constant sampling time (32 s, 
corresponding to a frequency span of 12.5 Hz). 
The data are fitted with a $1/\sqrt{N}$ dependence (dashed 
line), as well as with a constant offset term $\Delta \nu=a + b \sqrt{N}$ 
(continuous line). The second fit gives us an offset term 
$a=(5.9 \pm 2.4)$ mHz which is related to the systematics in 
data fitting and the stability of the fit parameters and the apparatus.  
On the right panel we report the error on the 
resonance curve fitting versus the number of averages, keeping 
constant the total measurement time in each of the four data 
sets, corresponding to 640 sec and its three next multiples. 
Each average corresponds to different FFT windows 
(spanning from 200 Hz in the case of 2 s sampling time to 3.125 Hz for 
128 s sampling time) and then different FFT frequency resolutions. 
Consistent with the plot on the left side, the peak frequency error 
on the right plot decreases from one data set to the other corresponding to 
increasing measurement time. The four curves also show a common 
trend with a minimization of the error for a number of averages 
between 8 and 16, corresponding to a frequency window $\delta \nu$ 
between 25 and 50 Hz, {\it i.e.} about 10 times the resonator 
linewidth of $\simeq 3$ Hz. Error bars represent the statistical uncertainty of the fitting procedure.}
\label{fig3}
\end{figure}

The thermal drifts are also an indirect reason for limiting the 
number of averages in a long-time measurement of the Casimir force. 
A measurement of the Casimir force around 3 $\mu$m 
capable of disentagling between the two different models for the temperature 
corrections, as discussed above and in \cite{Brown}, requires high force sensitivity. 
One possible detection scheme, which has been used to calibrate 
the apparatus, is based on the measurement of frequency shifts 
of the resonator induced by a spatially-dependent force \cite{Giessibl,Bressi1}
In the (ideal) case of a perfect stability (no drifts in 
frequency or other parameters), an ideal theoretical understanding 
of the response, and an infinite number of measurements, one could achieve  
an exact determination of the peak frequency for any value of the 
mechanical quality factor. Obviously, none of these conditions are 
fulfilled in an actual experiment. The parameters of the resonator may drift in 
time, there may be an incomplete understanding of the background which produces 
residuals in the fit, as well as statistical fluctuations 
due to the finite number of averages. This gives then rise to 
a finite precision in determining any parameter modeling the resonator 
response, including the peak frequency. We have studied the dependence 
of the precision in the determination of the peak frequency versus the number 
of averages of the FFT transform of the photodiode signal resulting from 
a fiber optic interferometer \cite{Rugar}. As shown in Fig. 3 (left) 
the error in the determination of the peak frequency is 
fitted by the standard $1/\sqrt{N}$ dependence. 
Due to the presence of thermal drifts, one cannot expect ideal behaviour 
even in the limit of infinite averaging. It is therefore important to 
study how to optimize the precision in the determination of the peak frequency for a 
fixed measurement time. Once the latter is fixed 
we can choose to measure on a large window (at the price of 
worse frequency resolution on the FFT analysis) or viceversa.
In Fig. 3 (right) we plot the error in the determination of the 
peak frequency versus the selected window of the FFT analyzer 
(or, equivalently, versus the sampling time for each measurement). 
It is evident that the error on the peak frequency is optimized 
for an intermediate value of the windowing. Larger windowing yields  
poor frequency resolution inherent to the FFT, partially compensated 
by the decreased statistical uncertainty resulting from the larger number of 
averages, while small windowing allows for better frequency 
resolution at the cost of large statistical fluctuations. Our tests 
indicate that the error on the peak frequency is minimized 
by using a window about ten times the intrinsic bandwidth 
of the mechanical resonator (on the order of 2-3 Hz in our case).
This also shows that, unless particular care is taken to 
minimize thermal drifts, the use of large mechanical quality 
factors for the resonators does not necessarily 
improve the sensitivity, as the longer measurement times required
 for the optimal narrower windowing will make the resonator more prone 
to thermal drifts: in the presence of  finite thermal stability 
there will be, for the frequency-shift measurement technique, an optimal 
quality factor for the resonator. 

\subsection{Parallel plane configuration: experimental issues}

\begin{figure}[b]
\setlength{\unitlength}{1cm}
\begin{center}
\includegraphics*[width=8.0cm,angle=0]{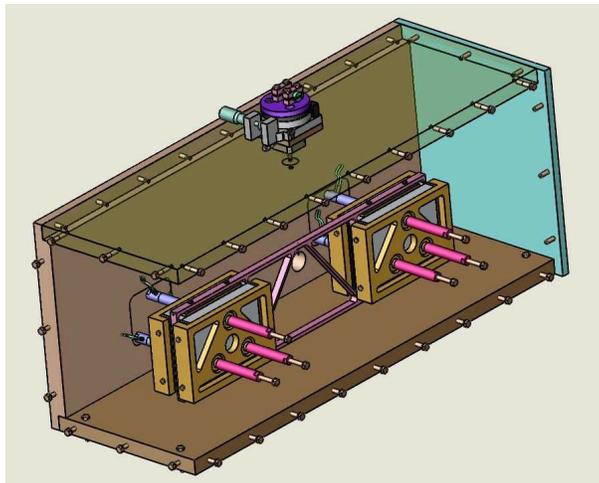}
\end{center}
\caption{Schematic design of the torsional balance under development 
at ILL, Grenoble. The parallelism between the plates 
of the balance and the outer plates is obtained through a set of 
high-precision piezoelectric actuators. The symmetry of the 
scheme allows for many cross-checks of systematic effects, and 
for physical calibrations and assessments of the ultimate 
sensitivity based on both electrostatic and gravitational forces, 
the latter force being testable through fine control of the 
position of nearby static masses.}
\label{fig4}
\end{figure}

The parallel plate configuration, as discussed in \cite{Onofrioproc}, has still 
a high priority for performing Casimir force measurements due to the unique 
features of this geometry. In particular, it may provide the strongest 
limits on non-Newtonian gravity in the 1-10 $\mu$m range due to the largest attainable
signal and the absence of gravitational force gradients apart from 
boundary effects \cite{Price,Onofrio,Carugno}. Due to the importance of 
investigating non-Newtonian forces in recent unification frameworks, this 
research program is currently pursued by our collaboration.
In order to measure the thermal contribution, the steep scaling of the parallel 
plane Casimir force with respect to distance requires the use of large, macroscopic surfaces, 
which in turn leads to the use of a high-sensitivity torsional balance \cite{Grenoble}. 
In our project, two pairs of plates are installed on opposite arms of the 
moving and static parts of the balance, respectively (see Fig. 4). 

These plates have a surface area of 120 cm$^2$ for 
the measurement at the largest gaps, on the order of 10 $\mu$m. 
For smaller distances ($\simeq$ 1$\mu$m) the effective area can 
be reduced to 15 cm$^2$, which minimizes the 
chance of having dust trapped in the gap. The target torque sensitivity 
is in the 1-100 $\mu$N/rad range, obtainable 
with quartz wire of diameter 50-150 $\mu$m. This allows for 
a minimum detectable force of order 1 pN. 
Three high precision piezoelectric actuators and their feedback 
controllers maintain a constant distance between the 
two plates throughout the duration of the measurement, with an 
accuracy of 0.2 $\mu$m. Capacitors symmetrically located on the 
opposite sides of the active surfaces for the measurement of 
the Casimir force allow for further control of the gap distance. 
The distance between the capacitor plates is large enough
 (100 $\mu$m) to avoid backaction effects on the 
torsional balance, yet provide adequate sensitivity. 
 
Parallelization procedures will be based on expertise already 
acquired in Grenoble through an experiment in which discretization 
of the energy levels of ultracold neutrons in the Earth gravitational 
field has been observed \cite{Nesvy}. While in these experiments parallelism 
between the neutron reflecting surface and the absorber 
has been kept to about 10$^{-6}$ radians, the expected parallelism 
with the use of the piezoelectric actuators and auxiliary capacitors 
is around 10$^{-7}$ radians. The surface roughness should be around 
1-2 nm. In a second stage of the experiment, metallic layers with 
thickness of $\simeq 1 \mu$m will be used when searching for 
hypothetical Yukawa forces with gravitational coupling, and this 
may lead to a degradation of the surface quality. The availability of 
{\it in situ} diagnostic techniques in surface science 
laboratories already present in Grenoble will provide 
complete characterization of the surfaces. 

\begin{figure}[t]
\setlength{\unitlength}{1cm}
\begin{center}
\includegraphics*[width=7.5cm,angle=0]{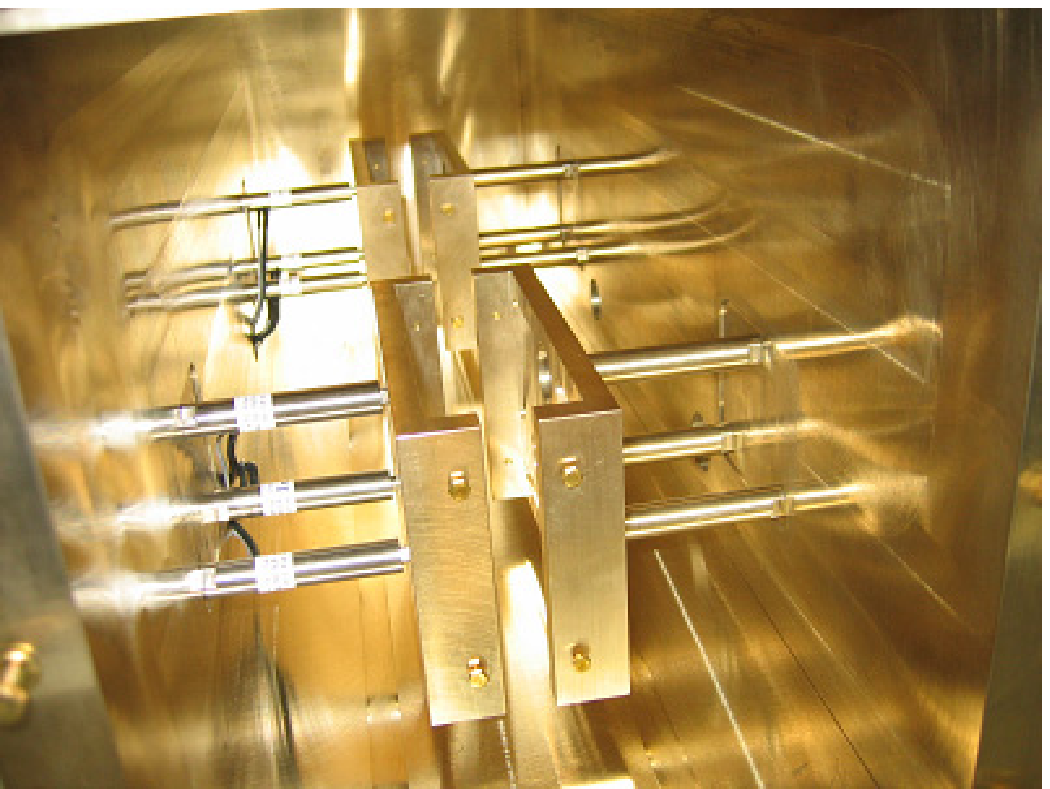}
\includegraphics*[width=7.5cm,angle=0]{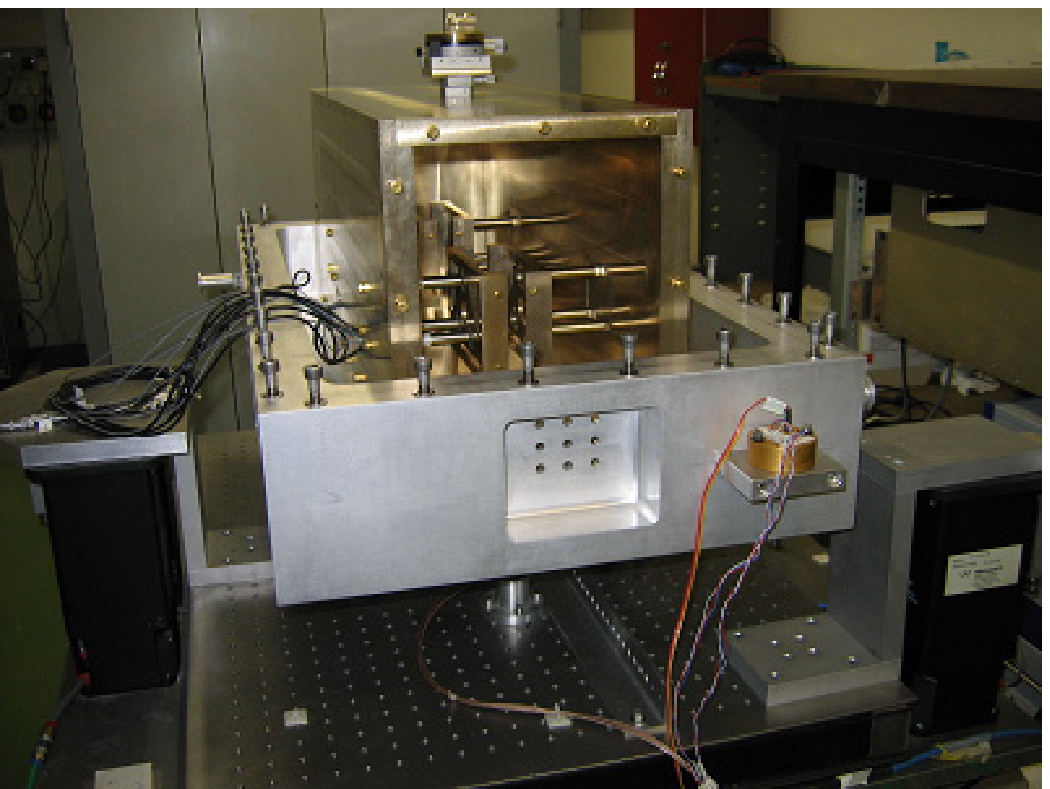}
\end{center}
\caption{Experimental set-up for the measurement of the Casimir 
force in a parallel plane configuration. On the left, the plate 
holders are connected to the external frame through three precision 
pieoelectric positioners each plate. 
On the right is an overall view of the vacuum chamber with the fiber holder on 
the top.}
\label{fig5}
\end{figure}

Another potential hurdle along the way is the tilting of the mirrors 
in the arms of the torsional balance. This can be addressed by 
reducing the lever of the torsional balance, properly designing 
the position of its pivot point, and segmenting the electrodes 
in such a way that small tilts do not result in changes of the 
effective surface. The latter solution also implies a control of 
the border effects, which have to be carefully estimated for 
instance through the worldline numerics approach pioneered in 
\cite{Gies}. The experiment is currently in preparation (see Figs. 5) and 
the first calibrations of the apparatus are planned in early 2006. 
Once again, we want to stress that various physical signals can  
be used to calibrate this apparatus, including 
the possibility of measuring the gravitational force through a 
Cavendish-like experiment. This allows for a  simple expression of the 
apparatus sensitivity in terms of the maximum distance at which 
a known gravitational source may be detected - a parametrization 
of the sensitivity particularly natural when discussing the 
limits to Yukawa forces in the micrometer range.

\section{Dissipative contribution to the Casimir force}

The emission of photons due to the dissipative nature of motion in
quantum vacuum has been predicted using various analytical 
\cite{Moore,Law,Lambrecht,Plunien,Dodonov,Crocce} and numerical 
\cite{Ruser}. Dissipation is expected when a material body undergoes 
a non uniform acceleration. The phenomenon is often discussed in the context of
a resonant cavity where the amplification of  vacuum field energy
is enhanced due to a periodic motion of a boundary and the
build-up of photons in the cavity. From this point of view the
phenomenon, also called dynamical Casimir effect, consists of the
amplification of vacuum field energy as the periodic modulation of
the cavity boundaries excites a particular mode of vacuum field,
with the transformed state being squeezed. 
\begin{figure}[b]
\setlength{\unitlength}{1cm}
\begin{center}
\includegraphics*[width=10.0cm,angle=0]{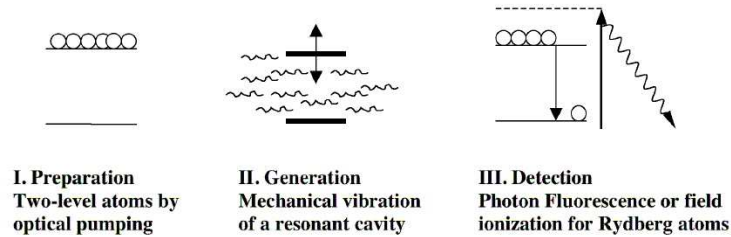}
\end{center}
\caption{Schematic of the atomic detection technique. 
The two-level atoms interact with the photons generated through the 
mechanical modulation of the cavity. 
In the detection stage, either fluorescence or field ionization 
techniques can be adapted to detect the change in the 
atomic population.}
\label{fig6}
\end{figure}
The number of photons
in the mode (if uncoupled  to any other mode) increases
exponentially as $ N_{cas} = \sinh^2(\lambda \epsilon t)$, where
$\epsilon=\delta L/L$ is the amplitude of the modulation and
$\lambda$ depends on the geometry of the cavity and on the
particular resonant mode \cite{jobcrocce}. In practice the
exponential growth is constrained due to the finite optical
quality factor $Q$, with the growth saturating at times 
$\tau_{sat} \simeq Q/\omega$. Typically, for cylindrical cavities with
rectangular or circular sections, $\lambda /\omega\sim {\cal
O}(1)$. 
Therefore the number of produced photons is approximated
as $N_{cas}\sim\sinh^2(Q\epsilon)$. For $\epsilon \simeq 10^{-8}$
(about a few nm displacement in the GHz range) and $Q \simeq
10^{8}$, a typical number of photons expected is on order of unity. 
The result is very sensitive to the value of $Q\epsilon$, the photon number 
 being $\simeq 10^3 $ for $Q\epsilon\sim 4$. These estimations illustrate the
difficulty of the observation of this phenomenon.
The number of created photons could be much larger if, instead of 
considering a cavity with moving boundaries, one considers a cavity with 
time-dependent reflectivity. The effective length of the cavity could be 
changed by irradiating with fast laser pulses a thin semiconductor film 
contained in it \cite{lozo1,lozo2, yablo}. 
It has been recently shown, using a simplified model \cite{crocce_pra}, that 
in this case one could reasonably get $Q\epsilon \gg 1$, but the heating due 
to the laser pulses and the generation of excitations in the semiconductor \cite{Dodo} 
are an open issue for the current implementation of this scheme as in the MIR 
experiment \cite{Braggio1,Braggio2,Braggio3}.
\begin{figure}[b]
\setlength{\unitlength}{1cm}
\begin{center}
\includegraphics*[width=13.0cm,angle=0]{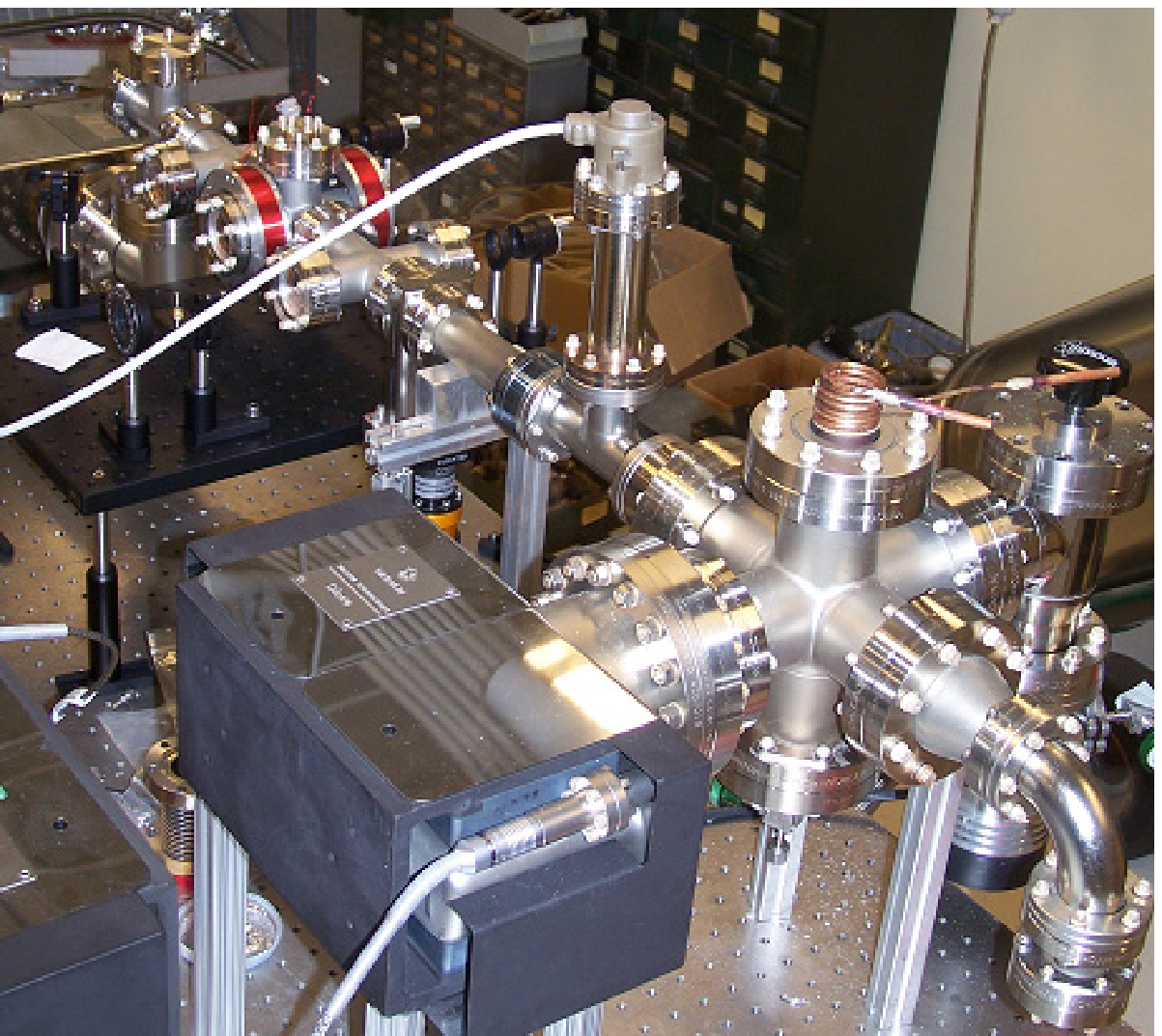}
\includegraphics*[width=13.0cm,angle=0]{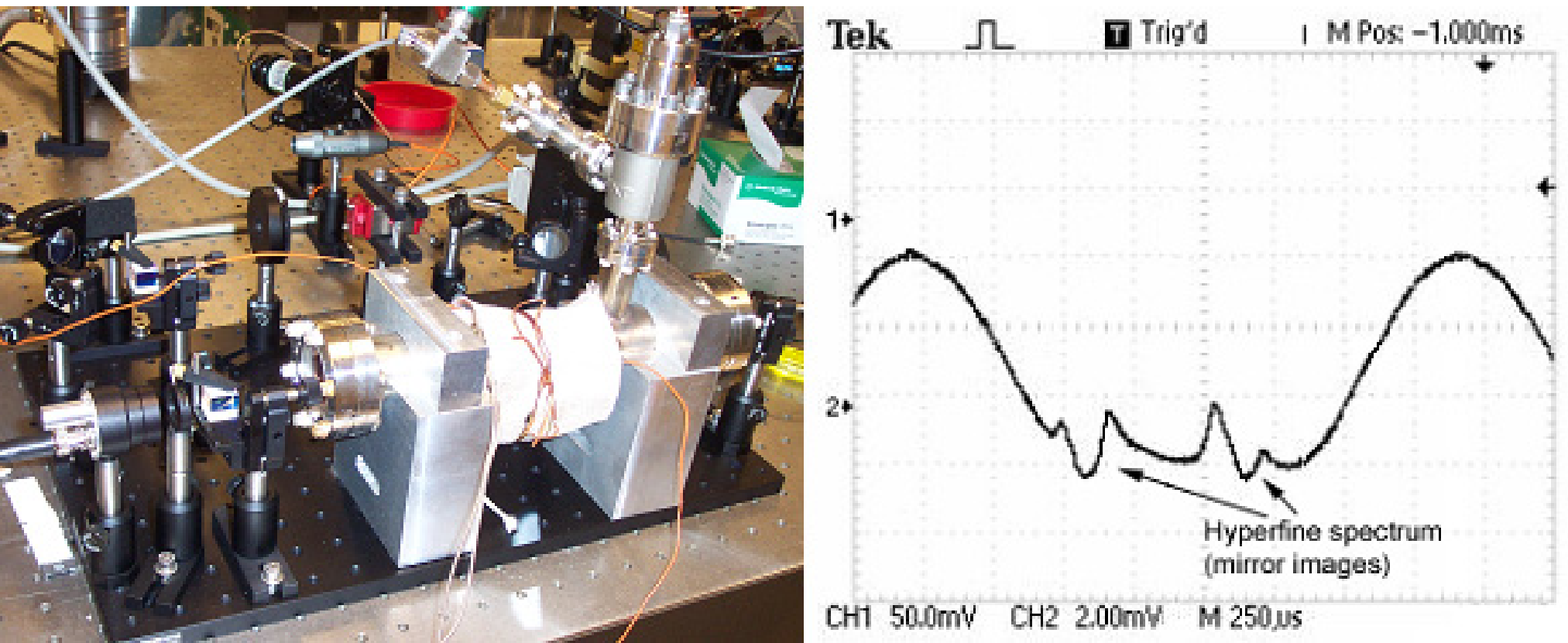}
\end{center}
\caption{Atomic beam line for studies of the induced transitions 
in hyperfine levels of $^6$Li by means of controllable electromagnetic 
fields (top picture).
$^6$ Li vapor cell (bottom left) and corresponding saturation absorption signal  
with the hyperfine states spaced by 228 MHz (bottom right).}
\label{fig7}
\end{figure}
Recently, a generation and detection scheme has been proposed based on 
the two emerging technologies of high-frequency resonators and the ultracold atoms 
\cite{Onofrioproc}. The basic idea is to prepare two-level atoms into a particular 
excited state and to induce the stimulated transition in the presence of the 
amplified vacuum field as schematically depicted in Fig. \ref{fig6}. 

The vacuum amplification takes place inside a resonant cavity where an array of mechanical 
resonators are coherently oscillating. These mechanical resonators of high frequency are 
available (up to 3.1 GHz) due to the recent development of the Film Bulk Acoustic Resonator 
(FBAR) devices. The integral part of the FBAR devices is a thin piezoelectric film made 
of Aluminum Nitride (AIN) for which the mechanical frequency can be changed with a varying thickness 
\cite{Ruby,Zhang}. Depending on the nature of two-level atoms, one 
could detect the progressive change in atomic population either by fluorescence 
(photon detection) or by field ionization procedure (electron detection).  
The original scheme relies on the hyperfine splitting of $^6\textup{Li}$ whose 
transition frequency in the ground state is 228 MHz, which is the lowest among 
alkali atoms, thereby allowing a reasonable matching with the working 
frequency of available high-frequency mechanical nanoresonators. 
Apart from the difficulty of fabricating an ensemble of mechanical resonators 
coherently driven at a well-defined frequency (456 MHz), the use of hyperfine 
splitting seems disadvantageous due to an extremely low stimulated emission rate. 
For a two-level atom, the stimulated emission rate is directly proportional 
to the spontaneous rate given by
\begin{equation}\label{eq:lifetime}
A=\frac {\omega^{3} D_{ij}^{2}}{\pi \epsilon_{o} \hbar c^{3}}
\end{equation}
where $D_{ij}^{2}$ is the matrix element for the dipole transition between \textit{i}th and 
\textit{j}th states. For the two hyperfine ground states of Li, the magnetic dipole interaction 
dominates the transition, which gives rise to an additional factor of $v^{2}/c^{2}$, lowering 
the emission rate by  $10^{20}$ compared to the usual electric dipole interaction. Furthermore, the estimated power of Casimir photons for
$Q\epsilon\sim 1$ is $N_{cas}\hbar \omega/\tau_{sat} \simeq
10^{-25} $W, too small to induce the hyperfine transition.
 
In order to maximize the atomic emission rate, one could consider an electric dipole transition 
with a larger matrix element. The most natural choice satisfying these criteria is  the use 
of circular Rydberg atoms \cite{Dodonov1,Dodonov2} having large electric dipole moment ($\simeq n^{4} a^{2}$) and 
relatively long lifetime ($\simeq n^{5}$) \cite{Haroche,Gallagher}, already employed in the studies of 
cavity quantum electrodynamics \cite{Hulet,Raimond} and quantum information processing\cite{Jaksch}.  
The transition frequency can in principle be chosen at will by varying the principal quantum 
number up to $n=150$ \cite{Yamada,Haseyama}, which provides a great tunability with a mechanical resonator. 

Although the use of Rydberg atoms greatly increases the stimulated emission rate, the 
number of photons generated inside the cavity is still too small to be measured through an appreciable 
change in atomic population. This issue can be overcome by exploiting an additional amplification 
procedure through super-radiance. 
The overall scheme is identical to the super-radiant maser system in the millimeter wave domain 
considered in \cite{RAM1,RAM2}. In super-radiance a single atom undergoing a transition 
triggers a subsequent emission on the neighboring atoms resulting in a large pulse whose 
lifetime is inversely proportional to the number of atoms present in the coherent volume. 
As long as the number of Casimir photons produced inside a resonant cavity exceeds the number 
of spontaneously emitted photons during the atomic travel time across the cavity, it is possible 
to measure the triggering effect of the Casimir photons on the super-radiant pulse through control of 
delay time, polarization, and phase \cite{RAM2}. Because the estimated number of Casimir photons is very small 
however, the triggering effect may be dominated by the vacuum fluctuations or the presence of 
thermal photons as well. For this super-radiance scheme to be effective, the Casimir photons 
must be the dominant triggering source. So far, we have built at Dartmouth a $^6$Li atomic 
beam source to study optical pumping and detection of hyperfine states populations (see Fig. 7), 
which can be easily extended to rubidium atoms, the most natural candidate for the 
preparation of Rydberg states in our scheme.

\section{Conclusions}
We have described ongoing experimental efforts in Casimir physics, with the goal of 
studying the interplay between {\it pure} quantum fluctuations and thermal 
effects, either in the form of thermal photons - always present in any finite 
temperature environment, or photons originating from the dissipative 
nature of non-uniformly accelerated motion of an object in quantum vacuum.
The combination of strong signals at large distances and relatively simple parallelization 
makes the investigation of the Casimir force in a cylinder-plane geometry quite attractive, 
allowing for discrimination between theoretical approaches. Recent technological 
advances make possible the achievement of almost ideal parallelization in the 
parallel plane geometry, providing strong constraints on hypothetical Yukawian forces.  
Use of hyperfine or super-radiant Rydberg atoms for the detection of the 
dynamical Casimir effect will require a delicate balance in the parameter space, 
and yet it promises exciting insights into the dissipative nature of quantum 
vacuum. These combined efforts should lead towards a more comprehensive understanding of 
fluctuation and dissipation of quantum vacuum.

\acknowledgments

This paper is based upon a plenary talk by R. Onofrio and a poster by W.J. Kim, who are grateful 
to the organizers of QFEXT'05 for the kind invitation and partial financial support, respectively.
M. Brown-Hayes acknowledges support from the Dartmouth Graduate Fellowship, W. J. Kim acknowledges 
support from Gordon Hull Fellowship program, and S.M. Middleman acknowledges support through an 
undergraduate Mellam fellowship. F. C. Lombardo and F. D. Mazzitelli acknowledge support from 
Universidad de Buenos Aires, CONICET and ANPCyT. We thank N. Monnig for experimental assistance, 
and R. L. Johnson for skillful technical support. We are grateful to M. Antezza, H. Gies, P. Milonni, 
V. Mostepanenko, L. Pitaevskii, and G. Ruoso for fruitful discussions.


\begin{thebibliography}{100}
\bibitem{note} The unique case of sphere-plane experiment using a lens as the spherical
surface gives a large signal \cite{Lamoreaux}, however the precision in the measurement
of the radius of curvature is an issue \cite{Lamoreauxerr}
\end{thebibliography}
\end{document}